# Brain structure can mediate/moderate the relationship of behavior to brain function and transcriptome: a preliminary study

Sumra Bari[1,†], Nicole L Vike[1,†], Khrystyna Stetsiv[2, %], Anne J Blood[3-5,%], Eric A Nauman[6%], Thomas M Talavage[6,%], Semyon Slobounov[7,&], Hans C Breiter[1,4,&]

[1]Department of Computer Science, University of Cincinnati, Cincinnati, OH, USA
[2]Warren Wright Adolescent Center Department of Psychiatry and Behavioral Sciences, Feinberg School of Medicine, Northwestern University, Chicago, IL, USA
[3]Mood and Motor Control Laboratory, Departments of Neurology and Psychiatry, Massachusetts General Hospital and Harvard Medical School, Boston, MA 02129, USA
[4]Laboratory of Neuroimaging and Genetics, Department of Psychiatry, Massachusetts General Hospital and Harvard Medical School, Boston, MA 02129, USA
[5]Martinos Center for Biomedical Imaging, Department of Radiology, Massachusetts General Hospital and Harvard Medical School, Boston, MA 02129, USA
[6]Department of Biomedical Engineering, University of Cincinnati, Cincinnati, OH, USA
[7]Department of Kinesiology, Pennsylvania State University, University Park, PA, USA

[†] indicate co-equal first authorship
[%] indicate co-equal second authorship
[&] indicate co-equal senior authorship

* *Correspondence:*

For project design, management, and data collection: Semyon Slobounov, 268P Recreation Building, The Pennsylvania State University, University Park, PA 16802, USA. Email: sms18@psu.edu

For hypotheses and conceptual framework, data analysis, and paper development: Hans C. Breiter, Department of Computer Science, College of Engineering and Applied Science, University of Cincinnati, 2901 Woodside Drive, Cincinnati, OH 45219, USA. Email: breitehs@ucmail.uc.edu

**Abstract**

Abnormalities in motor-control behavior, which have been associated with concussion and head acceleration events (HAE), can be quantified using virtual reality (VR) technologies. Motor-control behavior has been consistently mapped to the brain's somatomotor network (SM) using both structural (sMRI) and functional MRI (fMRI). However, no studies have integrated HAE, motor-control behavior, sMRI and fMRI measures. Here, brain networks important for motor-control were hypothesized to show changes in tractography-based diffusion weighted imaging [difference in fractional anisotropy (ΔFA)] and resting-state fMRI (rs-fMRI) measures in collegiate American football players across the season, and that these measures would relate to VR-based motor-control. We further tested if nine inflammation-related miRNAs were associated with behavior-structure-function variables. Using permutation-based mediation/moderation methods, we found that across-season ΔFA from the SM structural connectome (SM-ΔFA) mediated the relationship between across-season VR-based Sensory-motor Reactivity (ΔSR) and rs-fMRI SM *fingerprint similarity* ($p_{Sobel}^{perm}$ = 0.007 and $T_{eff} = 47\%$). The interaction between ΔSR and SM-ΔFA also predicted ($p_{F}^{perm}$= 0.036, $p_{\beta_3}^{perm}$= 0.058) across-season levels of ΔmiRNA-30d through permutation-based moderation analysis. These results suggest (1) that motor-control is in a feedback relationship with brain structure and function, (2) behavior-structure-function can be connected to HAE, and (3) behavior-structure might predict molecular biology measures.

## Introduction

Core features of both concussion and accumulated head acceleration events (HAE) include disturbances in balance, sensory-motor reactivity (SR), and spatial navigation – features quantifiable with virtual reality (VR) tasks[1–3]. Following Luria's framework[4], these three VR tasks can detect residual neurologic abnormalities in concussed athletes with cleared symptoms, unlike many computer-based neuropsychological tests[5] (see Supplemental Material for details and validation of the VR tasks).

Two of these three motor-control behavior processes (balance and SR) rely fundamentally on brain regions in the somatomotor resting-state network (SM). Somatomotor circuitry has been mapped extensively via structural MRI measures[6–13] and functional MRI (fMRI)- or PET-based assessments[14,15,24,16–23]. Brain regions in other resting-state networks, like subcortical (SUBC)[25,26], dorsal attention (DA)[27,28], and default mode network (DMN)[29], are also linked with motor-control behavior processes underlying balance and SR, whereas the fronto-parietal (FP)[30–32] and DA networks[32,33] are involved with spatial navigation (SN). The dearth of studies relating both structural and functional MRI to computational behavior (i.e., mathematical psychology and the advanced technology used to study behavior) measures of motor-control raises a question of what organizational framework the structure-function-behavior relationship might follow. Recent studies have constrained motor-control and other behaviors to be dependent variables (DV) in imaging[34,35]. Because it is now acknowledged that the adult brain is far more plastic than previously recognized[36–39], this raises the question whether changes in motor-control behavior might drive features of brain structure and/or function (i.e., that it may in some cases be the independent variable (IV) or mediator/moderator (Me/Mo) for structure-function-behavior relationships). The same question extends to whether any observation involving behavior and brain structure or function would predict a molecular biology metric such as micro-RNA (miRNA) levels, which have been reported to be elevated in football players and related to accumulated HAE[3,34,40,41]. The miRNA studied here have been implicated in multiple cellular processes, including those that follow the mechanical phase of impact, such as the initiation of inflammation and its longer-term resolution[42,43]. Accumulated HAE have been implicated in one of the mechanisms hypothesized to underlie brain injury, namely chronic inflammation which is energy intensive and can alter cellular processes[44–48]. Accumulated HAE have also been implicated in two other hypothesized mechanisms of brain injury: 1) alterations in cerebrovascular blood flow control[49–52], and 2) axonal abnormalities caused by traumatic shearing and stretching forces[53–56]. These models have been studied independently, and recently assessed together for relationships between markers of neuroinflammation and markers of cerebral blood flow[34].

The current study sought to connect markers of neuroinflammation with markers of structural abnormalities, as from axonal stretching, using a cohort of collegiate American football players. Prior work, using the same cohort of athletes, showed that across-season rs-fMRI *network fingerprint similarity* measures in motor-control networks (i.e., SM, DMN, SUBC, DA, and FP) were lower than in age-matched, non-athletes[40]. For this study we hypothesized that these athletes would exhibit across-season changes in resting-state fMRI (rs-fMRI) network connectivity (i.e., *fingerprint similarity*) in brain regions involved with motor-control, and that these measures would show a relationship with changes in diffusion tensor imaging (DTI) measures (i.e., changes in fractional anisotropy, $\Delta$FA) in the same brain networks. Further, it was hypothesized that these measures would be integrated with across-season changes in motor-control behavior (i.e., balance, SR, or spatial navigation measures). Our hypotheses further implemented permutation-based mediation (PMe)[57] and moderation (PMo)[58–60] to test if a panel of nine miRNAs (i.e., transcriptomic measures), previously implicated in inflammation and HAE[3], were associated with any two or three variables in a structure-function-behavior framework.

Given the focus on extending prior work to indirect measures of axonal abnormality, the ΔFA was always required in any three-way analysis involving structure, function, behavior, or miRNA variables with PMe/PMo. The use of permutation-based statistics with mediation/moderation analysis provides significant advantages for small-to-medium sized human subject cohorts, as detailed elsewhere[40,41] (see Supplemental Material). Permutation-based statistics with mediation/moderation analysis moves past standard association to identify more statistically mechanistic relationships. Unlike standard parametric statistics, permutation-based methods produce a true data distribution, thus protecting against false positives and increasing the overall power of the analysis. Variables involved in PMe and PMo were further analyzed against HAE measures collected during the football season. This integrated statistical approach applied across  structure, function, behavior and omic measures suggest a viable methodology for investigating complex multi-scale biology in humans, opening up avenues for bringing human studies in closer approximation to pre-clinical studies in animals. This approach can further identify potential biomarkers and expand our understanding of disease onset and progression.

## Methods

### Participants and data collection

Twenty-three male collegiate American football athletes (mean±s.d. = 21±2years) enrolled in this study. All participants provided informed consent, as approved by the Pennsylvania State University Institutional Review Board and in alignment with the Declaration of Helsinki. Participants were all seasoned starters who experienced a high frequency of HAE throughout the season (i.e., 16/23 athletes were non-speed linemen who sustain a high volume of impacts during practices and games)[117]. No participants were excluded due to history of concussion, or had a diagnosed concussion in the nine months preceding data collection. Two participants experienced concussion during the season but were clear to return to practices and games in one week. Only one participant was taken out of seasonal play due to a significant peripheral injury that occurred early on in the season; they were retained in the study cohort. Four participants reported family history of psychiatric illness but had no psychiatric illness themselves. Demographic information from each participant was confirmed by a team physician: race, age, years of play experience (YoE), player position, handedness, attention deficit hyperactivity disorder (ADHD) and dyslexia status, headache/migraine, and history of diagnosed concussion (HoC; reported as the number of previous concussions) (see **Supplemental Table 1)**. Three participants had a history of ADHD/dyslexia. Blood samples were collected at the beginning of the season prior to any contact practices (*Pre*) and within one week of the last regular season game (*Post*). Concurrent with blood collection, athletes also underwent virtual reality (VR) testing and MR imaging. HAE measures were collected over the season (i.e., between *Pre* and *Post* blood, imaging, and VR sampling) as described below. The same individuals were observed at both *Pre* and *Post*-season collection time points, and were continuously under the same dietary regimen, workout plan, and lifestyle management. *Pre*-season measures thus served as a strong comparison group for *Post*-season measurements, given no control group with similar body types, athletic plan, lifestyle management, and metabolic needs could be constructed.

**miRNA quantification**

Serum from *Pre* and *Post*-sessions were used to quantify miRNA levels as published before (Supplemental Material)[3]. Absolute levels of nine miRNAs as previously reported[3] (miR-20a, miR-505, miR-3623p, miR-30d, miR-92a, miR-486, miR-195, miR-93p, miR-151-5p) were integrated in statistical analysis.

**Resting-state fMRI (rs-fMRI)**

Athletes participated in two MRI sessions consisting of a ten-minute rs-fMRI scan, with the parameters as listed in Supplemental Material. rs-fMRI data were processed using in-house MATLAB code using AFNI and FSL functions as described in[118] and in Supplemental Material.

A functional connectivity matrix (i.e., the functional connectome; FC) ordered into seven cortical sub-networks, as proposed by Yeo et al.[119], and an additional eighth sub-network comprising sub-cortical regions (as defined in[120]) were computed. These networks included: Visual (VIS), Somato-Motor (SM), Dorsal Attention (DA), Ventral Attention (VA), Limbic System (L), Fronto-Parietal (FP), Default Mode Network (DMN) and subcortical regions (SUBC). The *fingerprint similarity* for FC and of these eight networks was calculated by correlating the submatrix corresponding to each of the networks from *Pre* and *Post*-sessions[121]. The network *fingerprint similarity* thus represents how close a participant's FC is between repeat visits.

## Diffusion Weighted Imaging and Fractional Anisotropy Analysis

Diffusion-weighted images were acquired at each imaging session with a spin-echo echo-planar imaging (EPI) sequence (11 min 28 s), with the parameters and the pre-processing pipeline to generate probabilistic tractograms as described in[122] and in Supplemental Material. Average fractional anisotropy (FA) was calculated across all streamlines for each brain parcellation unit (Supplemental Material) to generate a structural connectome (SC). Each SC matrix was ordered into eight cortical sub-networks as described for rs-fMRI. The across-season change in FA ($\Delta$FA) values for each network and SC were calculated by the average difference across the networks from *Pre* and *Post*-sessions and is thus represented as $\Delta$FA values.

## Virtual Reality (VR) based motor-control testing

Athletes completed a previously-validated VR testing using a 3D TV system (HeadRehab.com)[2,5,34,40,41] for *Pre* and *Post*-sessions. The test included scores from three modules: Spatial Memory, Sensory-Motor reactivity (SR, for whole-body reaction time), and Balance along with an overall Comprehensive score. To facilitate interpretation, all scores were scaled such that higher scores represent better performance[4]. See Supplemental Material for details.

## Head Acceleration Events (HAE) monitoring

HAE were monitored at all contact practice sessions (max = 53; no games were monitored as it could disturb competitive play) using the BodiTrak sensor system from The Head Health Network. Other studies have found the majority of head impacts occur during practices and not games[117]. Per athlete, HAE were quantified as the total number of sessions (sessions); cumulative number of hits exceeding the threshold *Th* = 25G and 80G ($cHAE_{25G}$ and $cHAE_{80G}$); average number of hits exceeding *Th* = 25G and 80G per session ($aHAE_{25G}$ and $aHAE_{80G}$) was normalized with total number of sessions (see Supplemental Material). 25G reflected routine impacts such as blocking, and 80G reflected play-ending tackles.

## Statistical Methods

To assess across-season changes, "Δ" values were calculated for all measures by subtracting *Pre*-session measures from *Post*. For rs-fMRI networks, the *fingerprint similarity* - representing the within-participant similarity - was used instead of Δ values. After imaging data quality assurance and missing data removal, 14-23 participants had both *Pre* and *Post* data for two or more measurements. All statistical analyses used the software package R. Analyses involved identification of two-way associations, discovery of their overlap as three-way associations, and mediation/moderation testing within a permutation-based framework. Secondary analyses involved testing of association between the four measures integrated in Me/Mo analyses against HAE measures.

### Two-way associations

Two-way associations were checked between nine network *fingerprint similarity* scores (FC, Vis, SM, DA, VA, L, FP, DMN and SUBC) with the corresponding network-ΔFA, four ΔVR scores (Balance, SR, Spatial Memory and Comprehensive), nine ΔmiRNA (miR-20a, miR-505, miR-3623p, miR-30d, miR-92a, miR-486, miR-92a, miR-93p, and miR-151-5p) and five HAE measures (sessions, $cHAE_{25G}$, $cHAE_{80G}$, $aHAE_{25G}$ and $aHAE_{80G}$). Linear regression was run between two variables and outliers were removed based on Cook's distance. After outlier removal, linear regressions were re-run and all two-way associations with $p \leq 0.05$ were passed forward for analysis with three-way associations. For all two-way associations meeting the $p \leq 0.05$ threshold, the number of outliers removed, $p$-value, β-coefficient and adjusted $R^2$ ($R^2_{adj}$) were reported.

**Three-way associations**

Two-way associations with $p \leq 0.05$ were used to build three-way associations. A three-way association between any three variables was formed if the two-way associations for the common variables were found to be significant, leading to the relation:

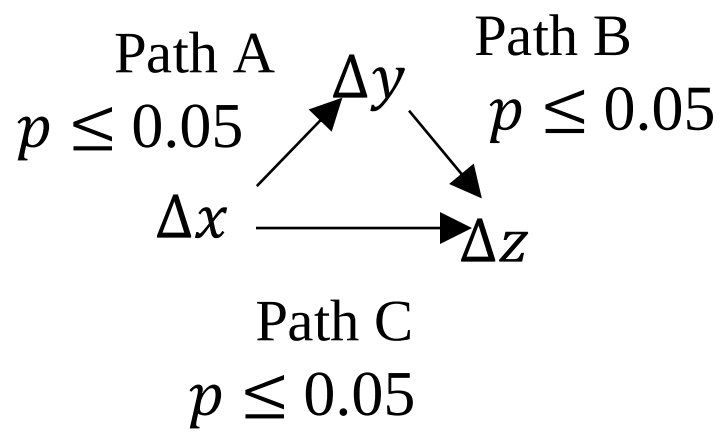

Here $\Delta x, \Delta y$ and $\Delta z$ are across-season measurements chosen from nine network *fingerprint similarity* scores, nine network-ΔFA values, four ΔVR scores and nine ΔmiRNA; *a priori* analyses focused on three constraints as detailed at the beginning of Results. The conjoint requirement of Path A, Path B, and Path C each meeting $p \leq 0.05$ is equivalent to a conjunction analysis threshold of $p \leq 0.000125$ or a Bonferroni correction for 400 tests (i.e., $p < 0.05/400$), which is more than the eligible set of three-way analyses. The eligible set of three-way analyses, given *a priori* constraints, was a maximum of 50 for ΔFA, rs-fMRI, and behavior associations, was a maximum of 45 for ΔFA, rs-fMRI, and miRNA associations, and was 90 for ΔFA, behavior, and miRNA associations.

**Moderation Analysis**

The moderation model proposes that the strength and direction of the relationship between independent variable (X) and dependent variable (Y) is controlled by the moderator variable (Mo). The moderation is characterized by the interaction term between IV and Mo in the linear regression equation as given below:

$$Y = \beta_0 + \beta_1 X + \beta_2 Mo + \beta_3 (X * Mo) + \epsilon$$

For each test, moderation is significant if $p_{\beta_3} \leq 0.05$ (the interaction term) and $p_F \leq 0.05$ (for the overall model). Here, $p_{\beta_3} \leq 0.05$ indicates that $\beta_3$ is significantly different than zero using a t-test and indicates a significant interaction between X and Mo predicts the value of the third variable (Y).

**Mediation Analysis**

Mediation was utilized to clarify the causal relationship between the independent variable (X) and dependent variable (Y) with the inclusion of a third mediator variable (M). The mediation model proposes that instead of a direct causal relationship between X and Y, the X influences M, which then influences Y. This statistical framework places M in the casual pathway between X and Y and moves far past standard association analysis[57,58]. In this study, mediation was required to have an effect > 30%.

**Permutation-based Mediation and Moderation Analysis**

Permutation-based tests are well-suited for studies with small to moderate sample sizes and also control for the occurrence of false positives as they only require the assumption that the two samples being compared are interchangeable in setting up the null hypothesis[59,60]. Unlike standard statistics, permutation-based methods produce a true distribution from the data under study, protecting against false positives and increasing power of the analysis.

We performed permutation-based mediation (PMe) and moderation analysis (PMo) as described in detail elsewhere, for all three-way associations (see Supplemental Material). Given prior work showing permutation frameworks are adequate to protect against false positives[61], no further correction for multiple comparisons was needed for PMe/PMo analyses.

## Results

Three layers of analysis were undertaken. First, all two-way associations meeting $p < 0.05$ were assessed and listed in **Supplemental Table 2(A-J)**. Then, three-way associations determined which of the four types of variables were submitted for PMe/PMo analyses. These three-way associations had three constraints: (i) ΔFA measures were required in the three-way relationship given the goal of testing neuroinflammation and axonal abnormality (e.g., ΔFA measures) relationships, (ii) the same network had to be in a relationship between the ΔFA measures and rs-fMRI *fingerprint similarity*, (iii) SM, DMN, SUBC, DA, and FP networks were an *a priori* focus for VR-based measures of balance, SR, and SN. For the second layer of analysis, variables in three-way associations were tested for PMe/PMo and were analyzed across all possibilities of being the IV, Me/Mo, and DV. The use of permutation statistics is a salient protection against false positives[61], but further corrections for multiple comparisons for the first and second layers of analysis were considered and detailed in Methods. As a third layer of analysis, variables observed in significant or trend PMe/PMo relationships were tested in two-way relationships with HAEs. Years of playing experience (YoE) and history of concussion (HoC) were initially considered as covariates for first layer analyses, but neither YoE or HoC showed significant associations with the variables considered for Me/Mo, and were not included.

### <u>Permutation-based Mediation Analyses</u>

PMe analysis showed that the SM-ΔFA variable mediated the relationship between variables for ΔSR and SM *fingerprint similarity* [**Figure 1**(A-C)] with $p_{Sobel}^{perm}$ = 0.007 and $T_{eff} =$ 47%. The pairwise interactions between the three variables (ΔSR, SM-ΔFA and SM *fingerprint similarity*) are depicted in **Figure 1**(B). There was a negative relationship between ΔSR and SM-ΔFA, a positive relationship between SM-ΔFA and SM *fingerprint similarity*, and a negative relationship between ΔSR and SM *fingerprint similarity*. As the mediator, SM-ΔFA showed a clear parametric relationship to ΔSR and SM *fingerprint similarity* (**Figure 1**(C)).

## Permutation-based Moderation Analyses

PMo analysis showed that the SM-ΔFA variable influenced the relationship between ΔSR and ΔmiR-30d variables [**Figure 2**(A-C)]. The SM-ΔFA variable (moderator) interacted with ΔSR (independent variable) to affect ΔmiR-30d (dependent variable) with $p_F^{perm}$ = 0.036, $p_{\beta_3}^{perm}$= 0.058. The pairwise interactions between the three variables (ΔSR, SM-ΔFA and ΔmiR-30d) are depicted in **Figure 2**(B). There was a negative relationship between ΔSR and SM-ΔFA, a positive relationship between SM-ΔFA and ΔmiR-30d, and a negative relationship between ΔSR and ΔmiR-30d. As the moderator, SM-ΔFA showed a clear parametric relationship to ΔSR and ΔmiR-30d (**Figure 2**(C)), emphasizing how, together with ΔSR, it moderates, and thus predicts, ΔmiR-30d.

To emphasize the commonality between the PMe and PMo results, **Figure 3** combines both results with HAE metrics (see below), confirming the relationship of these variables to head impacts across the football season.

## Relationship to HAE Variables

For variables observed in PMe/PMo relationships, further analyses found a negative relationship between ΔSR and $cHAE_{25G}$ (i.e., HAE above 25G related to increased time required to correct balance) and positive relationships between SM-ΔFA and sessions as well as SM *fingerprint similarity* and sessions (**Figure 3**). The relationship between HAE and other variables are listed in Supplemental Table 1.

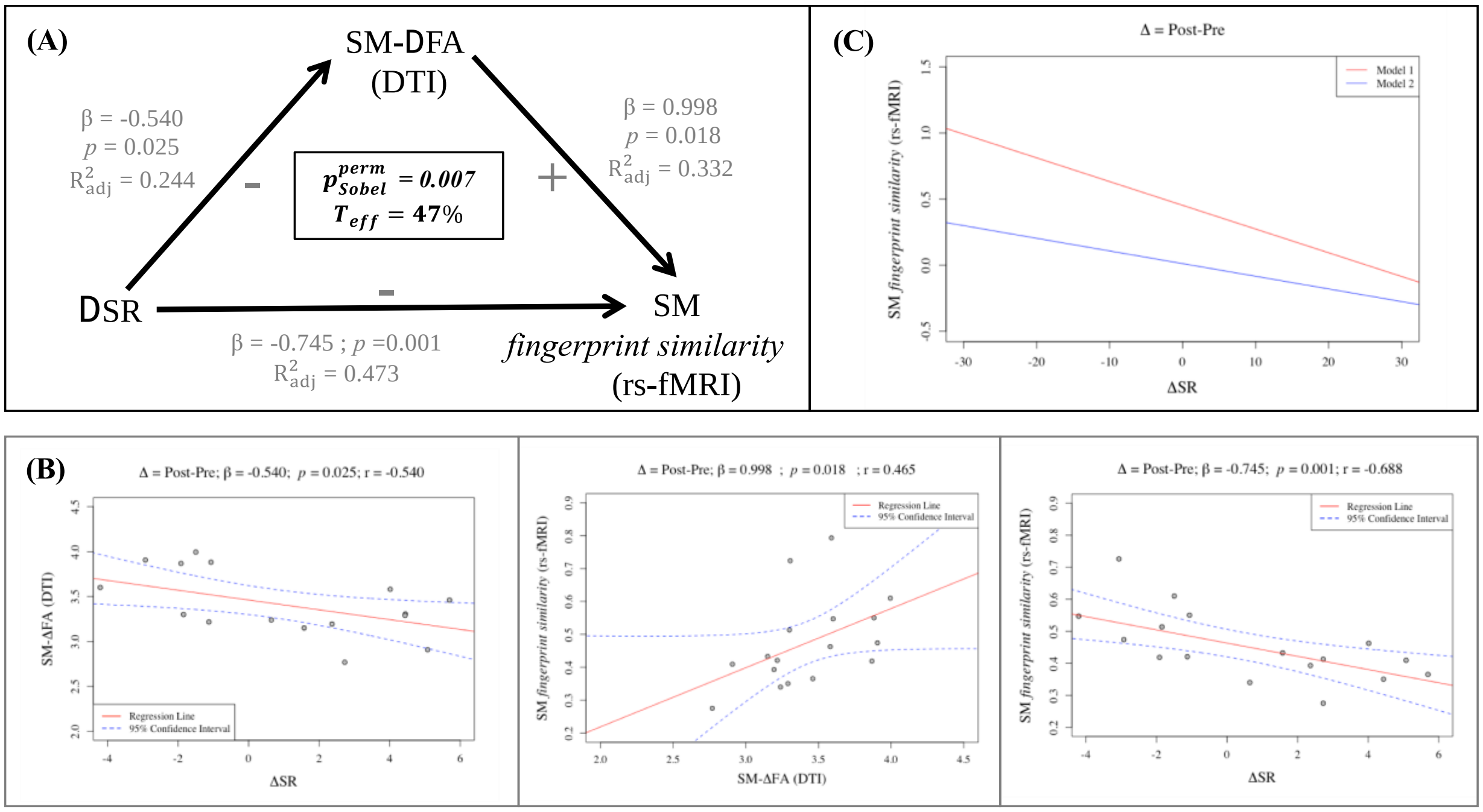

**Figure 1. (A-C)** Significant permutation based-mediation (PMe) result. The regression slope (β), *p*-value and adjusted $R^2$ ($R^2_{adj}$) depicted on each arm of the triangle corresponds to the significant two-way associations, after Cook's distance outliers removed, between those two variables with *p*-values reported at a significance level of 0.05. **(A)** SM *fingerprint similarity* was the dependent variable, ΔSR and SM-ΔFA were independent variable and mediator ( $p^{perm}_{Sobel} = 0.007, T_{eff} = 47\%$). **(B)** Two-way associations corresponding to the mediation analysis. There was negative relationship between ΔSR and SM-ΔFA; there was positive relationship between SM-ΔFA and SM *fingerprint similarity*; there was negative relationship between ΔSR and SM *fingerprint similarity*. **(C)** The plot depicts the change in slope between model 1, which plots the slope term for the regression between SM *fingerprint similarity* and ΔSR, and model 2, which plots the slope term for the regression between SM *fingerprint similarity* and ΔSR when SM-ΔFA was included in the regression model.

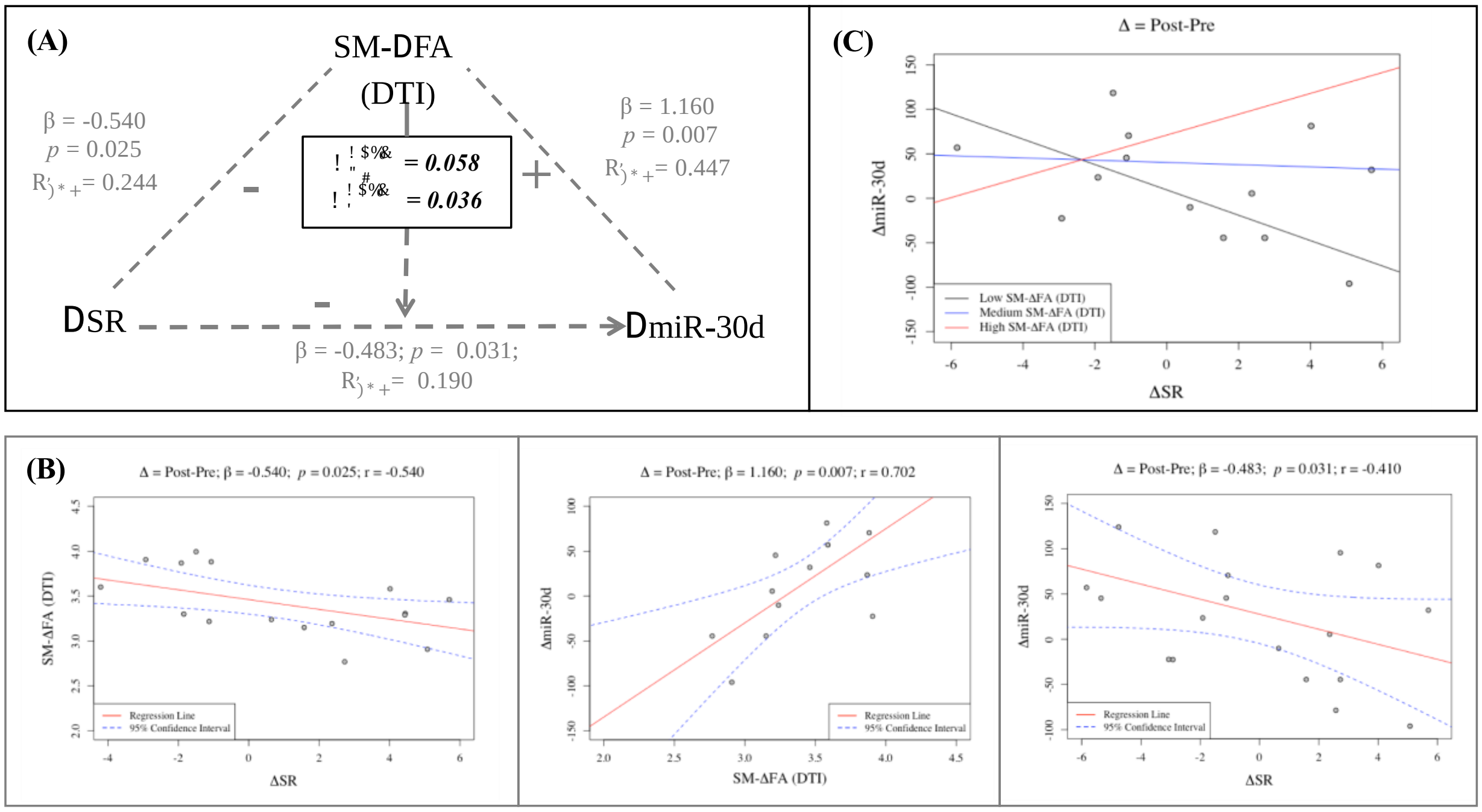

**Figure 2. (A-C)** Permutation-based moderation analysis (PMo) result. The regression slope (β), *p*-value and adjusted $R^2$ ($R^2_{adj}$) depicted on each arm of the triangle corresponds to the significant two-way associations, after Cook's distance outliers removed, between those two variables with *p*-values reported at a significance level of 0.05. **(A)** ΔmiR-30d was the dependent variable, ΔSR and SM-ΔFA were independent variable and moderator ( $p_F^{perm} = 0.058, p_{\beta_3}^{perm} = 0.036$). **(B)** Two-way associations corresponding to the moderation analysis. There was negative relationship between ΔSR and SM-ΔFA; there was positive relationship between SM-ΔFA and ΔmiR-30d; there was negative relationship between ΔSR and ΔmiR-30d. **(C)** The plot depicts the relationship between ΔSR and ΔmiR-30d with SM-ΔFA as the moderator. Three lines corresponds to low (mean - standard deviation), medium (mean) and high (mean + standard deviation) SM-ΔFA values.

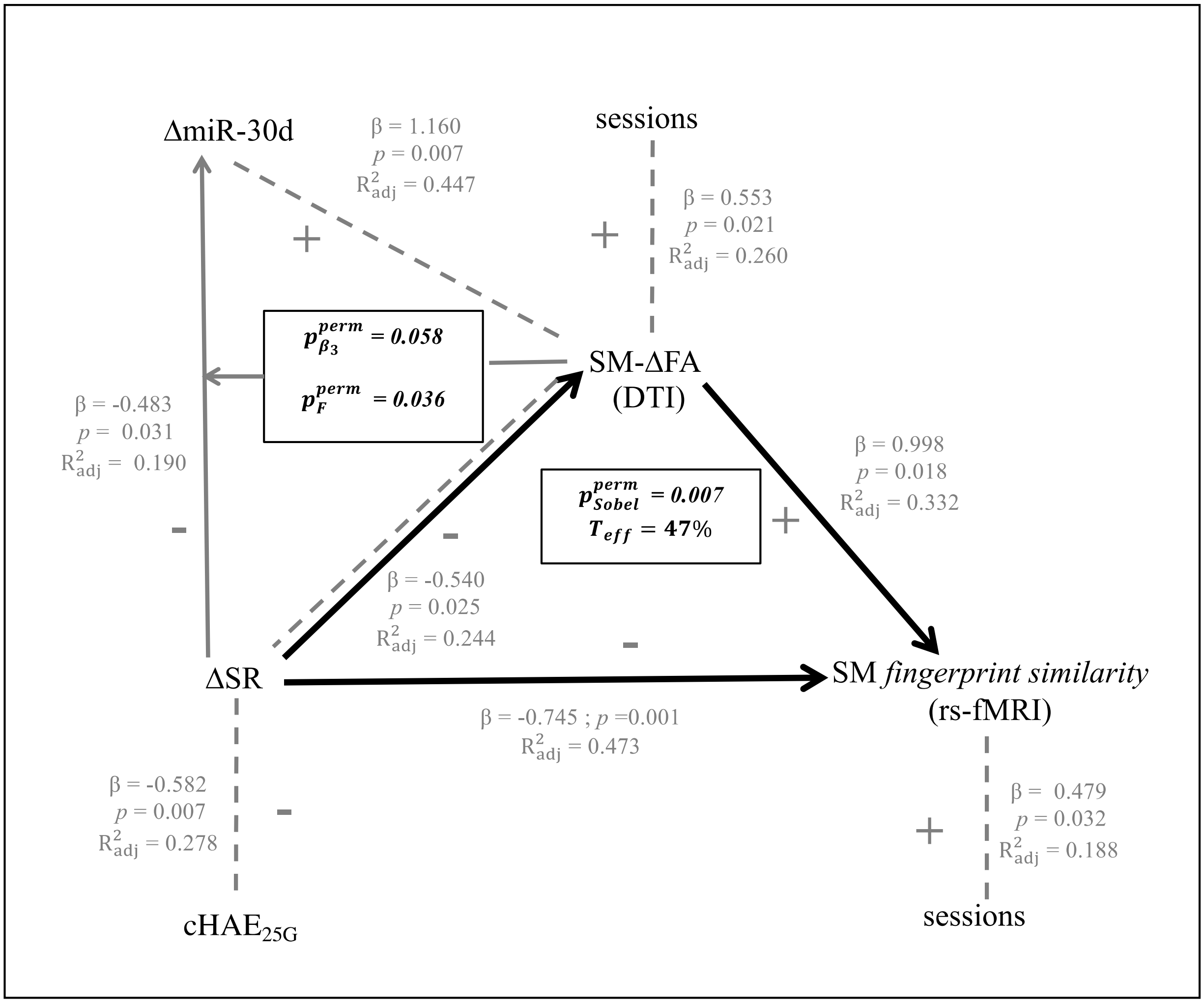

**Figure 3.** This figure combines the mediation and moderation analyses results with Head Acceleration Events (HAE). ΔSR was the independent variable, SM *fingerprint similarity* and ΔmiR-30d were the dependent variables and SM-ΔFA acted as the mediator and the moderator for the two results. The regression slope (β), *p*-value and adjusted $R^2$ ($R^2_{adj}$) depicted arm of the diamond corresponds to the significant interaction results, after Cook's distance outliers removed, between those two variables with *p*-values reported at a significance level of 0.05. ΔSR had a negative relationship with $cHAE_{25G}$ and SM-ΔFA and SM *fingerprint similarity* had positive relationships with sessions. This figure depicts the integration of resting state fMRI with diffusion weighted imaging (DWI), transcriptomics and computational virtual reality (VR) behavior task along with Head Acceleration Events (HAE).

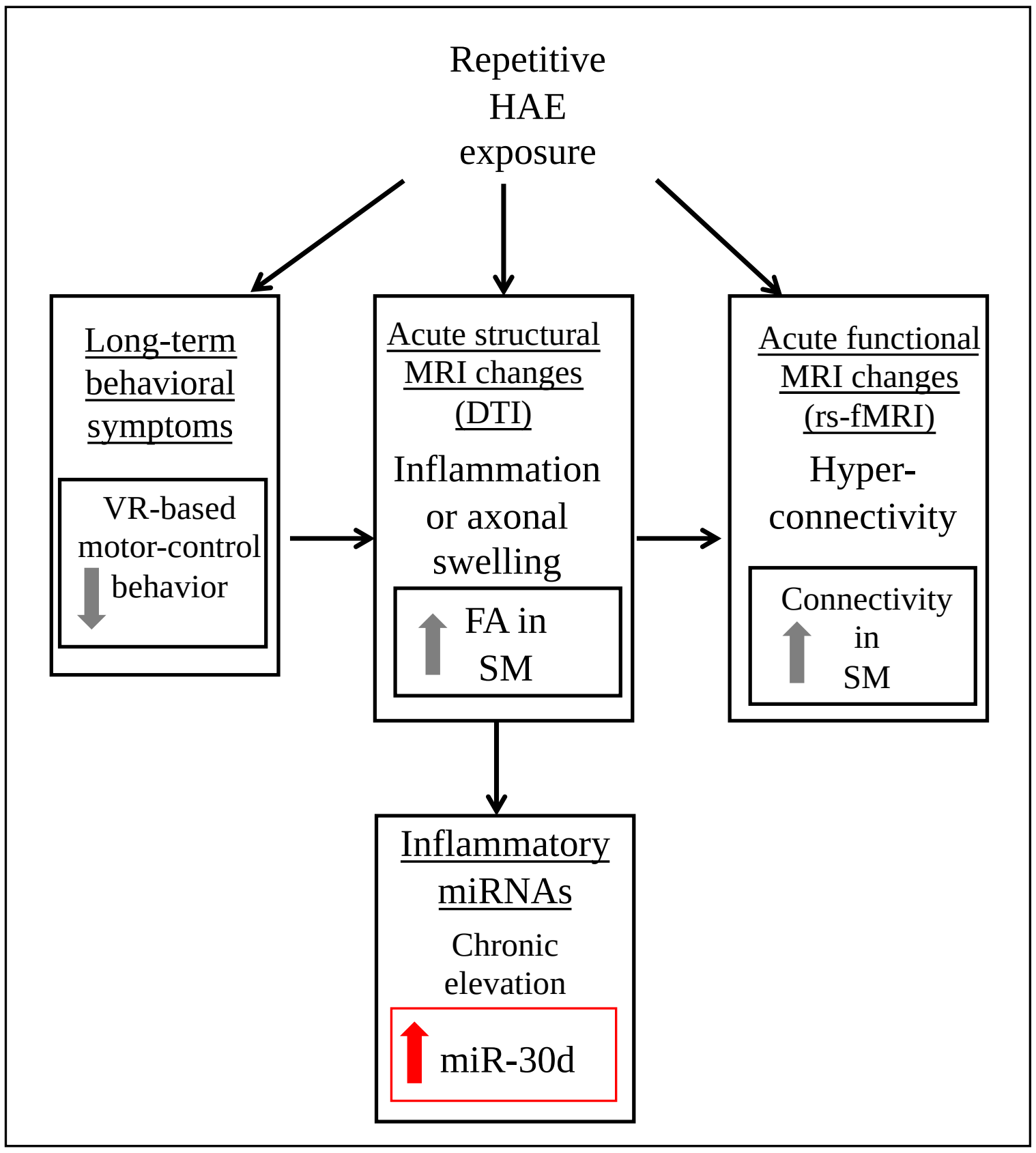

**Figure 4.** Synopsis figure summarizing the interplay between mediation and moderation variables. Repetitive head acceleration events (HAE) in football athletes have been related to changes in brain homeostasis, as evidenced by neuroimaging studies. In the present study, significant mediation and moderation effects were observed with ΔVR sensory-motor reactivity (SR) task performance, changes in FA in somatomotor network (SM-ΔFA), somatomotor network (SM) *fingerprint similarity*, and ΔmiR-30d. MiR-30d, inflammatory miRNA, significantly increased from *Pre* to *Post*-season, whereas increasing trend effects were observed in FA and rs-fMRI connectivity in SM and decreasing performance in SR from *Pre* to *Post*. Previous neuroimaging studies have associated increase in FA with inflammation or axonal swelling due to accumulated HAE. Hyper-connectivity in rs-fMRI networks have been associated with brain's short-term compensatory mechanism to use multiple alternative pathways to process information when primary pathways have been damaged due to HAE. In this study behavior variable have a feedback relationship with brain and molecular biology variables pointing to how brain structure, brain function, and behavior can be related and connected to the effects of HAEs in contact athletics, and predict molecular biology measures.

**Table 1.** Permutation-based mediation result. The primary mediation hypothesis assumed that ΔSR was the independent variable (X), SM-ΔFA was the mediator (M), and SM *fingerprint similarity* score was the dependent variable (Y). Secondary mediation was run with SM-ΔFA as the X, ΔSR as the M, and SM *fingerprint similarity* score as the Y. The regression slope (β), $p$-value after Cook's distance outliers removed and permutation-based Sobel $p$-value ($p_{Sobel}^{perm}$) and total effect mediated ($T_{eff}$; expressed as %) are reported. Results with $p_{Sobel}^{perm} < 0.05$ and $T_{eff}$ % $\geq 30$ were considered significant.

| **Mediation Variables** | | | | **Step 1: X predicting Y** | | **Step 2: X predicting M** | | **Step3: M predicting Y** | | **Step 3: X (with M) predicting Y** | | **$T_{eff}$** | **$p_{Sobel}^{perm}$** | N |
|---|---|---|---|---|---|---|---|---|---|---|---|---|---|---|
| **Y** | **Mediation** | **X** | **M** | **c** | **$p_c$** | **a** | **$p_a$** | **b'** | **$p_{b'}$** | **c'** | **$p_{c'}$** | **%** | | |
| **SM *fingerprint similarity*** | ***Primary*** | **ΔSR** | **SM-ΔFA** | -0.745 | 0.001 | -0.540 | 0.025 | 0.998 | 0.018 | -0.334 | 0.162 | 47.00 | 0.007 | 15 |
| | Secondary | SM-ΔFA | ΔSR | 0.998 | 0.018 | -0.540 | 0.025 | -0.745 | 0.001 | 0.525 | 0.037 | 26.00 | 0.050 | 15 |

**Table 2.** Permutation-based moderation result. The regression slope (β), $p$-value after Cook's distance outliers removed and permutation-based $p$-values ($p_{F}^{perm}, p_{\beta_3}^{perm}$) for moderation are reported. X is the independent variable, Mo is the moderator and Y is the dependent variable in moderation analysis.

| **X** | **Mo** | **Y** | **Path A: X predicting Mo** | | **Path B: Mo predicting Y** | | **Path C: X predicting Y** | | **$p_{\beta_3}^{perm}$** | **$p_{F}^{perm}$** |
|---|---|---|---|---|---|---|---|---|---|---|
| | | | **Std $\beta$** | **$p_1$** | **Std $\beta$** | **$p_2$** | **Std $\beta$** | **$p_3$** | | |
| ΔSR | SM-ΔFA | ΔmiR-30d | -0.540 | 0.025 | 1.160 | 0.007 | -0.483 | 0.032 | 0.058 | 0.036 |

**Discussion**

This preliminary study analyzed the directional relationships between DTI- and rs-fMRI-associated brain networks, motor-control behavior measures, and inflammatory miRNAs linked with HAEs and concussion . Variables in directional relationships were further assessed against HAEs. Unlike prior multimodal imaging-omics work[34,35,40], no assumptions were made regarding which variables could be IVs, Me/Mos, or DVs in PMe/PMo analyses. All variables submitted to PMe/PMo were required to show (a) two-way associations meeting a nominal significance threshold ($p<0.05$), (b) three-way relationships between sets of two-way associations., and (c) a minimum Sobel percent effect based on prior studies[33,34,40,41]. When setting up three-way associations for PMe/PMo, three *a priori* constraints were imposed: (i) ΔFA measures needed to be in the three-way relationship, (ii) networks assessed between DTI and rs-fMRI needed to be the same (e.g., *fingerprint similarity* for rs-fMRI and across-season change in ΔFA were observed in the same networks), and (iii) the SM, DMN, SUBC, DA, and FP networks were an *a priori* focus for analysis with VR-based measures for balance, SR, and SN. We observed a significant PMe relationship where (1) the across-season change in ΔFA for the SM network (SM-ΔFA) mediated the relationship between (2) across-season change in SR (ΔSR; speed associated with correcting balance perturbations) and (3) the rs-fMRI *fingerprint similarity* for the same SM network (SM *fingerprint similarity*) (**Figure 1**). Namely, a structural brain measure (ΔFA) sat in the causal pathway between a motor-control variable and a functional brain measure (rs-fMRI). The same across-season changes in SR and DTI for the SM network also produced a trend effect for moderating across-season change in an inflammatory miRNA (miR-30d) (**Figure 2**). In particular, the interaction between a change in motor-control behavior and change in brain structure predicted the change in a miRNA level. Together these PMe/PMo findings suggest a model of feedback effects across behavior and brain structure, function, and neuroinflammatory control (**Figure 4**).

Several features of these Me/Mo relationships are unique and should be discussed in terms of the current literature.

First, the independent variable in both Me/Mo relationships was SR, which quantifies the time it takes a participant to correct their stance when there are perturbations to their balance (**Figure 3**). In the context of mediation, the across-season change in SR was related to a change in brain function (rs-fMRI *fingerprint*), and this relationship was mediated by a change in brain structure (SM-ΔFA). To our knowledge, no head impact study has shown a relationship between behavior, brain structure, and brain function. Furthermore, we find no reports showing that a relationship between brain structure and function for the *same* brain region/network links to a motor-control behavior known to be connected to that region/network. The directional implications of this observation are relevant in that no other conformations of these three variables produced significant effects. Motor-control behavior clearly affected synchrony of the resting-state fMRI signal across the SM network through its effects on the DTI signal in the SM network. A standard brain mapping assumption is that behavior is the output of the brain; in this case, however, behavior was the input and thus must be classified as a feedback signal[62–65]. The field of plasticity has long recognized that behavior itself changes synaptic strength and organization[66]. There is also an emerging literature about healthy control behavior and patient treatment consistent with the idea that white matter plasticity occurs in adults, and not only during development[67,68,77–84,69–76]. Our findings are consistent with this literature and suggest the possibility that acute motor-control impairment due to subconcussive events may itself drive brain changes (**Figure 4**) and lead to longer term impairment that impedes complete recovery.

A parallel point must be made regarding the trend moderation result, where the interaction between ΔSR and across-season ΔFA supports a hypothesis that they can *predict* the change in a transcriptomic measure, specifically a miRNA with known relationships to inflammatory processes (**Figures 2,3**). This observation also can only be interpreted as a feedback relationship where behavior affects a molecular biology variable (miR-30d). The miRNA panel tested in this study has known roles in inflammation and inflammatory diseases[85–88], cancer progression[89–98], and neurodegeneration[99–103]. In the same cohort of athletes, miR-30d has been previously implicated in regional brain perfusion changes in the basal ganglia[34] and changes in energy-related metabolites[41], grounding the current findings in clear neurophysiological abnormalities in brain regions important for motor-control. Behavioral changes alone are rarely observed and difficult to replicate in studies of contact athletes without diagnosed concussion[104,105]. Although research points to significant neuroimaging and biochemical changes in this population across a sports season[105,106], behavior is not commonly shown to correlate with these changes[104,105]. As observed with the mediation result, an across-season increase in the time with which participants corrected balance perturbations (i.e., SR) impacted the level of an inflammatory miRNA.

All three variables from the PMe analysis, and two of same variables from the PMo analysis, showed significant relationships to HAE: (a) a higher number of practices increased the change in the two imaging measures, and (b) low level HAE had a negative effect on behavior (**Figure 3**). Previous neuroimaging research has demonstrated the role head impacts play in neurophysiological changes, such as altered connectivity (rs-fMRI and fMRI), decreased brain volumes, neurochemical alterations (MRS), altered CBF (perfusion imaging), and axonal injury (DWI)[34,48,105,107,108]. Accumulated HAE were related to an increase in SR score (i.e., increased time to adjust posture in the direction of an altered virtual environment), increased miR-30d levels, increased average FA across the SM network, and a an increase in the average rs-fMRI connectivity in the SM network at *Post*-season (**Figure 4** and **Supplemental Figure 1**). Hyper-connectivity in rs-fMRI DMN[109,110] and increased FA values[111,112] as observed at *Post*-season have been previously observed in high school football athletes just after the season ended. Hyper-connectivity in rs-fMRI networks have been associated with the brain's short-term compensatory mechanism to use multiple alternative pathways to process information when primary pathways have been damaged due to HAE[109]. The increase in FA, due to mild traumatic brain injury, has been associated with axonal swelling or inflammation leading to constricted axonal pathways, thereby reducing perpendicular diffusion and increasing FA[113–116]. In this study, the number of sessions was associated with a change in both structural *and* functional imaging variables for the SM network, and HAE were associated with motor-control behavior.

Several limitations must be noted for this study. First, there were no intermediate sample collection time points during the season for any behavior or imaging measure. Blood collections and VR tests were only conducted twice – once prior to contact practice (*Pre*-season) and once directly following the end of the regular competitive season (*Post*-season). To observe more transient changes in these metrics, or more complex dynamics in the structure of change, it would be beneficial to collect data at more time points during and after the season. Another unavoidable limitation was that the HAE metrics were only collected at practices, and not during competitive play (as per policies to not disturb game preparation and play), which resulted in a less complete picture of the true exposure to HAE. Second, this study lacked age-matched, non-contact athlete controls, although individual baseline measures taken before the season served as appropriate comparison group for *Post*-season measures given these porticipants were strictly monitored for diet, exercise, and lifestyle (e.g., grades, extra-curricular activities were managed). It is difficult to match these variables between athletes who participate in different sports and it could require a potentially overwhelming array of covariates. A third caveat to this work is that it only studied male participants; future studies need to also incorporate females and include more sophisticated metrics of sex steroid cycles (e.g.[35]).

A fourth caveat needs to be acknowledged, namely that the sample size was moderate. To account for this limitation, a rigorous permutation-based statistical approach was applied while also controlling for false positives. Permutation based analysis, unlike standard statistics, produces a true distribution controlling for false positives and increasing the overall power of the analysis. Permutation methods only require the assumption that the two samples being compared are interchangeable in setting up the null hypothesis, as opposed to the many that are required for standard parametric statistics; the violation of any of these assumptions leads to higher false positives. Collection of detailed multi-modal data in humans can lead to smaller samples, which is more likely to violate the central limit theorem if inference testing does not follow a permutation framework[60]. Permutation-based statistics, which permute the actual data, produce a meaningful distribution for exact inference[59].

## Conclusion

In summary, this preliminary study of American collegiate football athletes confirmed the hypothesis that brain networks involved with motor-control [e.g., the somatomotor network (SM)] would show relationships between DTI (SM-ΔFA) and resting-state fMRI (SM *fingerprint similarity*) measures, and would be statistically integrated with changes in motor-control behavior (i.e., SR). We further found that the interaction of SR and DTI signal across-season predicted, at a trend level, the change in an inflammatory miRNA (miR-30d) previously linked to brain perfusion, energy-related metabolites, and behavior[34,41]. These two sets of results strongly suggest behavior can be in a feedback relationship with brain and molecular biology variables. The integration of these multi-scale biological measures through a permutation-based approach uncovered a potential mechanism of brain alterations due to HAE as summarized in **Figure 4**. Broadly, these results point to how brain structure, brain function, and behavior can (1) be related, (2) connected to the effects of HAEs in contact athletics, and (3) predict molecular biology measures.

**Acknowledgements**: We thank the Penn State football players for their effort participating in this study, along with Katie Finelli and Madeleine Scaramuzzo for their assistance with subject recruitment and data collection. We would like to thank Dr. Zoran Martinovich and Dr. Jaroslaw Harezlak for providing guidance for the statistical analysis.

**Declarations**

**Funding:** All work in this paper was funded by the listed academic institutions, and without specific NIH, NSF, or DoD support. Data collection was funded by SS's lab. Funding for sMRI and rs-fMRI analysis, data integration, and manuscript preparation was funded by HB's lab (Warren Wright Adolescent Center).

**Conflicts of interest/Competing interests:** The authors declare they have no financial or other conflicts of interest with regard to the data and analyses presented herein.

The opinions expressed herein are those of the authors and are not necessarily representative of those from their respective institutions.

**Ethics approval:** This study was approved by Pennsylvania State University Institutional Review Board, in alignment with the Declaration of Helsinki.

**Consent to participate:** All participants provided informed consent, as approved by the Pennsylvania State University Institutional Review Board.

**Consent for publication:** All authors have contributed to this manuscript and consent to its review and publication.

**Availability of data and materials:** All data used in analyses will be deposited in a public repository selected by Brain Imaging and Behavior, in anonymized format.

**Code availability:** All codes used in analyses will be deposited in a public repository selected by Brain Imaging and Behavior, in anonymized format.

**Author contributions:** Study concept/design: SB, NLV, HCB; Acquisition of original data: SS; Coding of statistical tools: SB and NLV (with guidance from TMT, EAN, HCB); Analysis of data: SB and NLV (with guidance from HCB), Interpretation of data: SB, NLV, HCB (with significant guidance from AJB, along with input from KS, TMT, EAN, SS); Statistical assessment: SB and NLV (with guidance from HCB); Authored original draft: SB, NLV, HCB; Generated figures: SB; Revision of manuscript for content: SB, NLV, KS, AJB, EAN, TMT, SS, HCB; All authors approved the final version of the paper for submission.

**Supplemental Material**

**Methods**

## Participants and data collection

Twenty-three male collegiate varsity football athletes (mean± standard deviation = 21±2 years) enrolled in this study. The same individuals at both *Pre* and *Post*-season collection time points were used who were under the same dietary regimen, workout plan, and lifestyle management at the two time points. *Pre*-season measures thus served as a strong comparison group for *Post*-season measurements, given no control group with similar body types, athletic plan, lifestyle management, and metabolic needs could be constructed. All participants provided informed consent, as approved by the Pennsylvania State University Institutional Review Board and in alignment with the Declaration of Helsinki. Four participants reported family history of psychiatric illness but had no psychiatric illness themselves. Demographic information from each participant was confirmed by a team physician: race, age, years of play experience (YoE), player position, handedness, attention deficit hyperactivity disorder (ADHD) and dyslexia status, headache/migraine, and history of diagnosed concussion (HoC; reported as the number of previous concussions) (see **Supplemental Table 1)**. Three participants had a history of ADHD/dyslexia. Blood samples were taken at the beginning the football season prior to any contact practices (*Pre*) and within one week of the last regular season game (*Post*). Blood samples were prepared and sent out for miRNA quantification. Concurrent with blood collection, athletes also underwent virtual reality (VR) testing and MR imaging sessions. HAE measures were collected over the season (i.e., between *Pre* and *Post* blood, imaging, and VR sampling) as described below. Due to inadequate levels of serum for miRNA quantification in a small number of participants, the number of participants decreased from 23 to between 18-20 depending on the miRNA of interest. One athlete did not participate in any MR imaging sessions and two did not participate in *Post* MR imaging session leading to 20 participants with complete MR imaging data. As a result of the missing data the final cohort consisted of 14-23 participants with two or measures described below at *Pre* and *Post*.

**Serum Extraction**

Five mL of venous blood were drawn from each participant at *Pre* and *Post* sessions. Samples were placed in a serum separator tube, allowed to clot at room temperature, and then centrifuged. Serum was extracted from each tube and pipetted into bar-coded aliquot tubes. Serum samples were stored at -70°C until they were transported to 1) a central laboratory for blinded miRNA batch analysis[1,2] and 2) Metabolon (Morrisville, NC, USA) for blinded metabolite analysis.

**miRNA quantification**

Serum samples collected at *Pre* and *Post* sessions were used to isolate and quantify levels of RNA. 100 $\mu$L of serum was aliquoted and RNA was isolated using a serum/plasma isolation kit (Qiagen Inc., Venlo, Netherlands) as per the manufacturer's protocol. RNA was eluted in 20 $\mu$L of DNAse/RNAse-free water and stored at -80°C until further use.

Droplet digital PCR (ddPCR; Bio-Rad Inc., Hercules, CA, USA) was used to quantify absolute levels of nine miRNA (miR-20a, miR-505, miR-3623p, miR-30d, miR-92a, miR-486, miR-195, miR-93p, miR-151-5p)[1]. Prior to ddPCR analysis, RNA was checked for quality using a bioanalyzer assay with a small RNA assay. After quality confirmation, 10 ng of RNA was reverse transcribed using specific miRNA TaqMan assays as per the manufacturer's protocol (Thermo Fisher Scientific Inc., Waltham, MA, USA). Following the protocol as detailed in[1]. The final PCR product was analyzed using a droplet reader (Bio-Rad Inc., Hercules, CA, USA). Total positive and negative droplets were quantified, and from this, the concentration of miRNA/$\mu$L of the PCR reaction was reported. All reactions were performed in duplicate.

**Resting state fMRI (rs-fMRI)**

Athletes participated in two MRI sessions consisting of a 10-min eyes-closed rs-fMRI scan using echo-planar imaging with the following parameters: echo time (TE) = 35.8 ms, repetition time (TR) = 2000 ms, flip angle = 90°, 72 contiguous 2-mm axial slices in an interleaved order, voxel resolution = 2 mm × 2 mm × 2 mm, matrix size = 104 × 104 and 300 total volumes. One high-resolution $T_1$ scan using 3D magnetization prepared rapid acquisition gradient recalled echo (3D MPRAGE) sequence was acquired for registration and tissue segmentation purposes with the following parameters: TE = 1.77 ms, time of inversion (TI) = 850ms, TR = 1700ms, flip angle = 9°, matrix size = 320×260×176, voxel size=1mm×1mm×1mm, receiver bandwidth = 300 Hz/pixel, and parallel acceleration factor = 2.

rs-fMRI data were processed using functions from AFNI and FSL and using in-house MATLAB code for the processes detailed below. rs-fMRI BOLD timeseries were processed in the subject's native space and the first four volumes were discarded to remove spin history effects. Structural $T_1$ images were denoised and segmented into gray matter (GM), white matter (WM) and cerebrospinal fluid (CSF) tissue masks. The 4D BOLD timeseries was then passed through outlier detection, despiking, slice timing correction, volume registration, aligned to the $T_1$ structural scan, voxel-wise spatial smoothing within tissue masks, scaled to a maximum (absolute value) of 200, and the data were censored to remove outlier timepoints (with the censoring criteria as in[3]).

The timeseries were then detrended using no global signal regression with the following common regressors: (1) very low frequency fluctuations as derived from a bandpass [0.002–0.01

Hz] filter, (2) the six motion parameters and their derivatives, and (3) the voxel-wise local neighborhood (40mm) mean WM timeseries.

For connectivity analysis on a regional basis, the grey matter brain atlas[4] was warped to each subject's native space by linear and non-linear registration. This brain parcellation consists of 278 regions of interest (ROIs). Note that data from the cerebellum (comprising a total of 30 ROIs) were discarded, because the acquired data did not completely cover this structure for all participants. This resulted in a final GM partition of 248 ROIs. A functional connectivity matrix (namely the functional connectome; FC) was computed for each rs-fMRI scan through correlation of the mean (pre-processed) time series from each of the 248 ROIs. The resulting square, symmetric FC matrices were not thresholded or binarized. The *fingerprint similarity* for FC and these eight networks was calculated by correlating the submatrix corresponding to each of the networks from *Pre* and *Post* sessions[5]. The network *fingerprint similarity* thus represents how close a participant's FC is between repeat visits.

**Diffusion Weighted Imaging**

For each imaging session diffusion-weighted images were acquired with a spin-echo echo-planar imaging (EPI) sequence (11 min 28 s) with the following parameters: 72 contiguous $2\ mm$ axial slices in an interleaved order, matrix size = $110 \times 110$, voxel size = $2mm \times 2mm \times 2mm$, $TE = 94ms$, $TR = 9.8s$, 30 diffusion-weighted volumes (one per gradient direction) with $b = 1000 s/mm^2$, 30 diffusion-weighted volumes with $b = 2000 s/mm^2$, seven volumes with $b = 0$, and parallel imaging acceleration factor $= 2$. Only the 30 diffusion-weighted volumes with $b = 1000 s/mm^2$, and the volume with $b = 0$ that was acquired with these diffusion-weighted volumes were selected to calculate the diffusion metrics.

The DWI data were processed following the MRtrix3 guidelines (http://mrtrix.readthedocs.io/en/latest/tutorials/hcp_connectome.html) and the steps listed by Amico et al. 2018 [6] in an in-house build MATLAB code. In summary, a tissue-segmented image appropriate for anatomically constrained tractography was generated; multishell, multitissue response function was estimated (MRtrix command *dwi2response*) and performed the multishell, multitissue constrained spherical deconvolution (MRtrix *dwi2fod*); afterward the initial tractogram (MRtrix command *tckgen*, 10 million streamilines, maximum tract length = 250, FA cutoff = 0.06) was generated. Average fractional anisotropy (FA) was calculated across all streamlines for each of the 248 ROIs (brain parcellation as described above for rs-fMRI) to generate a structural connectome (SC). Averaged FA was calculated across all the streamlines for each of the 248 ROIs to generated a structural connectome (SC). The resulting square, symmetric SC matrices for each of the DWI scans were not thresholded or binarized. Each SC matrix was ordered into eight cortical sub-networks as described for rs-fMRI. The ΔFA values for each of the network and SC were calculated by the average difference of each of the networks from *Pre* and *Post* sessions and is thus represented as network-ΔFA values.

**<u>Virtual Reality (VR) based motor-control testing</u>**

VR testing was first described and validated by Slobounov et al., 2006[7] and Teel et al. (2015, 2016)[8,9]. The software used to display the virtual reality animations was developed and provided by HeadRehab, LLC (Chicago, IL). The HeadRehab Performance Test Software allows for the use of a range of interactive devices and depending on the system used, a variety of options are viable. These different systems have all been validated against one another. Overall, the subject will interact with the HeadRehab SideLine Performance Test Software modules via interactive devices or motion tracking devices. This technology was used to create “moving room” experiments with two conditions: (1) the virtual room could move as a whole structure or (2) separate components of the room could move in isolation (e.g., only the front wall moves).

In this study, a 3DTV system was used. A laptop was connected to the TV with an HDMI cord and the Sideline v8.1 test TV program was used. Display resolution was set at 1920 x 1080 (high definition, 1080p) with a 16:9 aspect ratio for use with the 3DTV system. Graphics without Stereo Effect were also used. A VCube head-tracking device and MoBar interactive device (in this case an Xbox controller) were used for interaction with the 3DTV. The VCube uses acceleration and translation to track the subject's head movement. The MoBar allows the subject to navigate and interact with the software. The program has a subprogram built in (Start Device Manager) that is required to be run to ensure that the devices are connected and functioning properly.

Once the athlete was properly set up, they underwent three virtual reality tasks: 1) balance, 2) Sensory-Motor reactivity (SR, for whole-body reaction time), and 3) spatial memory. The scores from each test were normalized and combined to produce a comprehensive score. These tests were based off initial findings from Dr. Alexander Luria and have been validated to detect functional abnormalities following mild traumatic brain injury[7,10] .

The Balance module tested an athlete's ability to maintain posture with a changing virtual environment (reported and validated by Teel et al., 2015[8]). Each visual scene was 30 seconds in duration and athletes were instructed to hold a tandem Romberg position for each trial (Romberg, 1853). For the first baseline trial, athletes were instructed to remain as still as possible while there were no changes to the visual field. The next six trials consisted of changes to the virtual environment in one of three directions (yaw, pitch, or roll) via two methods: (1) whole-room forward-backward oscillations with 18 cm displacement at 0.3 Hz or (2) whole-room lateral rolls at 10-30 degrees and 0.3 Hz. Deviances from baseline in the yaw, pitch, and roll directions, were recorded and interpolated (scale 0-10) such that a higher score indicated better task performance (10 being the best).

The SR module tested the time it took for each athlete to adjust their posture in the direction of an altered virtual environment. Athletes stood shoulder-width apart with their hands on their hips and were instructed to move their body with the direction of the changing virtual environment (along the anterior-posterior axis at 0.2 Hz). Randomly, the room would shift to shifts in the medial-lateral axis. When this occurred, athletes were asked to bend at the waist in the same direction as the virtual environment (i.e., bend left or right). Both the time and direction of the room shifts were randomized. Response times, in milliseconds, to the abrupt changes in the virtual environment were retrieved using accelerometers and results were reported as latency to shift the body. Five total trials were completed, with the first trial being a practice trial and the remaining four trials being used in the score calculation. The measured SR and errors in anticipation were calculated, interpolated, and concerted into a whole-body SR score ranging from 0 to 10. Higher scores were indicative of better performance.

The Spatial Memory (virtual corridor) module tested an athlete's ability to recall a virtual pathway. Athletes were shown a randomized virtual pathway (to avoid practice effect, which in fact was documented and published in Slobounov & Sebastianelli[11] with multiple turns leading to a door, followed by a return trip. Then, athletes were asked to repeat the pathway from memory using the remote. It should be noted that athletes were allowed to practice using the remote prior to any trials. If the athlete navigated the path correctly, testing was finished. If the athlete navigated the path incorrectly, the computer would replay the desired route and the athlete was given three total chances to correctly navigate. Outcomes were reported as (1) the average time of task completion and (2) the number of errors made. A score between 0-10 (10 being the best) was calculated based on the average time a participant stayed on the correct path. For each error made, a 3.33-point reduction was applied. If the athlete was unsuccessful after three attempts, their score was zero.

The Comprehensive score was calculated by combining the three module scores (Balance, SR, and Spatial Memory) into a ten-point scale (0 = worst, 10 = best).

**Head Acceleration Events (HAE) monitoring**

Head acceleration events (HAE) were monitored at all contact practice sessions (max = 53; no games were monitored) using the BodiTrak sensor system from The Head Health Network. It should be noted that other studies have found the majority of head impacts occurred during practices and not games [12]. Sensors were individually mounted on the inner surface, between the shell and the padding, of each active player's helmet using 3M VHB adhesive prior to contact practices beginning. Sensors were monitored throughout the season for integrity and functionality and outputs included peak translational acceleration (PTA; G-units) and impact location. The two G-unit thresholds (*Th*), ≥ 25G to < 80G (*Th* = 25G) and ≥80G (*Th* = 80G) used in this study were selected based on both pilot data and previous reports of impacts related to brain health and injury [13]. For each athlete the HAE were quantified as cumulative number of hits exceeding the threshold *Th* = 25G and 80G ($cHAE_{25G}$ and $cHAE_{80G}$):

$$cHAE_{Th} = \sum_{k=1}^{N} u(PTA_k - Th) \qquad (1)$$

where *u(x)* is the step function

$$u(x) = \begin{cases} 1 \; if \; x > 0 \\ 0 \; if \; x \; \leq 0 \end{cases}$$

The average number of hits exceeding *Th* = 25G and 80G per session for each athlete ($aHAE_{25G}$ and $aHAE_{80G}$) was normalized with total number of sessions

$$aHAE_{Th} = \frac{cHAE_{Th}}{total \; number \; of \; sessions} \quad (2)$$

**Permutation Testing**

To control for the occurrence of false positives due to multiple hypotheses testing, permutation-based moderation analysis was conducted. Permutation tests re-sample observations from the original data multiple times to build empirical estimates of the null distribution for the test statistic being studied [14,15]. Permutation-based tests are especially well-suited for studies with small to moderate sample sizes as they estimate the statistical significance directly from the data being analyzed rather than making assumptions about the underlying distribution. Permutation statistics do not require: (i) random selection of samples, (ii) sample independence, (iii) Gaussian distributions, (iv) sample sizes large enough for the central limit theorem to work, or (v) similar variance in samples being compared. Absent these assumptions, traditional statistical analyses produce more false positives and statistical power decreases. Permutation methods only require the assumption that the two samples being compared are interchangeable in setting up the null hypothesis [14,15] and, unlike standard statistics, permutation-based methods produces a true distribution from the data under study, thus increasing power of the analysis. To perform permutation testing, the test statistic is first obtained from the original data set, then the data is randomly permuted multiple (S) times and the test statistic is computed on each permutated data set. The statistical significance is computed by counting (K) the number of times the statistic value obtained in the original data set was more extreme than the statistic value obtained from the permuted data sets, and dividing that value by the number of random permutations (K/S) [14,15].

**Permutation-based Moderation Analysis (PMo)**

The moderation model proposes that the strength and direction of the relationship between independent variable (X) and dependent variable (Y) is controlled by the moderator variable (Mo). The moderation is characterized by the interaction term between IV and Mo in the linear regression equation as given below:

$$Y = \beta_0 + \beta_1 X + \beta_2 Mo + \beta_3 (X * Mo) + \epsilon$$

Moderation is significant if $p_{\beta_3} \leq 0.05$ and $p_F \leq 0.05$, where $p_{\beta_3} \leq 0.05$ indicates that $\beta_3$ is significantly different than zero using a t-test and indicates a significant interaction between IV and Mo.

For this study, permutation-based moderation analysis was performed for all three-way associations following the steps listed below:

1. Moderation analysis was performed by assigning the original data variables $\Delta\boldsymbol{x_i}, \Delta\boldsymbol{y_j}, \Delta\boldsymbol{z_k}$ as IV, DV and Mo to obtain reference test-statistics: $t_0$ and $F_0$. Only variables that formed three-way associations were considered.

2. Data permutation: values were randomly sampled without replacement from $\boldsymbol{x}_{\mathbf{1},\boldsymbol{i}}$ and $\boldsymbol{x}_{\mathbf{2},\boldsymbol{i}}$ to assign to $\boldsymbol{x}'_{\mathbf{1},\boldsymbol{i}}$ and $\boldsymbol{x}'_{\mathbf{2},\boldsymbol{i}}$.

3. Across season measures were computed from the permuted dataset $\Delta\boldsymbol{x}'_{\boldsymbol{i}} = \boldsymbol{x}'_{\mathbf{2},\boldsymbol{i}} - \boldsymbol{x}'_{\mathbf{1},\boldsymbol{i}}$. Similarly, $\Delta\boldsymbol{y}'_{\boldsymbol{j}}$ and $\Delta\boldsymbol{z}'_{\boldsymbol{k}}$ were computed.

4. Moderation analysis was performed on the permuted dataset $\Delta\boldsymbol{x}'_{\boldsymbol{i}}$, $\Delta\boldsymbol{y}'_{\boldsymbol{j}}$, $\Delta\boldsymbol{z}'_{\boldsymbol{k}}$ by assigning as IV, DV and Mo and the test statistics: $t'_q$ and $F'_q$ were obtained.

5. The counter variable $K_1$ was incremented by one if absolute value of $t_0$ was greater than absolute value of $t'_q$.

6. $K_2$ was incremented by one if absolute value of $F_0$ was greater than absolute value of $F'_q$.

7. Steps 2-6 were repeated: $q = 1,2, \cdots, S$ times. Here, $S = 100{,}000$.

8. Permutation-based $p$-value $p_{\beta_3}^{perm}$ was calculated as the proportion of the $t'_q$ values that are as extreme or more extreme than $t_0$ i.e. $K_1/S$ .

9. Permutation-based $p$-value $p_F^{perm}$ was computed from $F_0$ and $F'_q$ i.e. $K_2/S$.

10. Moderation analysis was considered significant if $p_{\beta_3}^{perm} \leq 0.05$ and $p_F^{perm} \leq 0.05$.

**<u>Permutation-based Mediation Analysis (PMe)</u>**

Mediation was utilized to clarify the causal relationship between the independent variable (X) and dependent variable (Y) with the inclusion of a third mediator variable (M). The mediation model proposes that instead of a direct causal relationship between X and Y, the X influences M, which then influences Y. Beta coefficients and their standard error (s) terms from the following linear regression equations, following the four step process of Baron and Kenny (1986)[16,17], were used to calculate Sobel $p$-values and mediation effect percentages (Teff):

$$Step\ 1 : Y = \gamma_1 + c(X) + \epsilon_1$$

$$Step\ 2 : M = \gamma_2 + a(X) + \epsilon_2$$

$$Step\ 3 : Y = \gamma_3 + c'(X) + b(M) + \epsilon_3$$

Step 4: Sobel's test was then used to test if $c'$ was significantly lower than $c$ using the following equation:

$$Sobel\ z - score = \frac{c - c'}{\sqrt{b^2 s_b^2 + a^2 s_a^2}} = \frac{ab}{\sqrt{b^2 s_b^2 + a^2 s_a^2}} \qquad (3)$$

Using a standard 2-tail z-score table, the Sobel $p$-value was determined from the Sobel z-score and the mediation effect percentage ($T_{eff}$) was calculated using the following equation:

$$T_{eff} = 100 * \left[1 - \frac{c'}{c}\right] \qquad (4)$$

For this study, permutation-based mediation analysis was performed for only variables that formed three-way relationships following the steps listed below:

1. The non-permuted across-season data variables $\Delta\boldsymbol{x_i}, \Delta\boldsymbol{y_j}, \Delta\boldsymbol{m_k}$ were assigned as independent, dependent, and mediator variables, respectively. Linear regressions $\Delta x_i \rightarrow$ $\Delta y_j$, $\Delta x_i \rightarrow \Delta m_k$ and , $\Delta m_k \rightarrow \Delta y_j$ were run to obtain Cook's outliers from each regression. All outliers indicated by the three regressions were removed to obtain a common set of participants.

2. A four-step process for mediation analysis [17,18] was run on non-permutated data variables with all outliers removed to obtain the reference Sobel z-score ($z_0$).
3. Data permutation: values were randomly sampled without replacement from $\boldsymbol{x}_{\mathbf{1},i}$ and $\boldsymbol{x}_{\mathbf{2},i}$ to assign to $\boldsymbol{x}'_{\mathbf{1},i}$ and $\boldsymbol{x}'_{\mathbf{2},i}$.
4. Across-season measures were computed from the permuted dataset $\Delta\boldsymbol{x}'_i = \boldsymbol{x}'_{\mathbf{2},i} - \boldsymbol{x}'_{\mathbf{1},i}$..
5. Similarly, $\Delta\boldsymbol{y}'_j$ and $\Delta\boldsymbol{m}'_k$ were computed.
6. Mediation analysis following steps 1 through 2, as above, was performed on the permuted dataset $\Delta\boldsymbol{x}'_i$, $\Delta\boldsymbol{y}'_j$, $\Delta\boldsymbol{m}'_k$ and the test statistic ($z'_s$) was obtained.
7. The counter variable $K$ was incremented by one if the absolute value of $z_0$ was greater than the absolute value of $z'_s$.
8. Steps 2-6 were repeated: $s = 1,2, \cdots, S$ times. Here, $S = 100{,}000$.
9. Permutation-based Sobel $p$-value ($p^{\text{perm}}_{\text{Sobel}}$) was calculated as the proportion of $z'_s$ values that were as extreme or more extreme than $z_0$ (i.e., $K/S$ ).

Secondary mediation analysis is done by flipping the independent and mediator variables and repeating the entire above process. Secondary mediation is run to confirm the directed causal pathway between independent, mediator and dependent variables.

Primary mediation results were considered significant if $p$-values associated with terms a, b, and c (terms derived from unpermuted data) were < 0.05; c' ≥ 0.05 and $p^{perm}_{Sobel} < 0.05$, $T_{eff}$ was > 30%, and $p^{perm}_{Sobel} \geq 0.05$, $T_{eff} \leq 30\%$ for the secondary mediation.

**Supplementary Tables and Figures**

**Supplemental Table 1**. Demographic information obtained from each participant: Race (W = White, AA= African American), Age (years), Years of play experience (YoE), history of diagnosed concussion (HoC; reported as the number of previous concussions), player Position, Handidness, Headache/Migraine and ADHD and Dyslexia status (here indicated under the ADHD column).

| ID | Race | Age | YoE | HoC (#) | Position | Handidness | Headache/Migraine | ADHD |
|---|---|---|---|---|---|---|---|---|
| 1 | W | 20 | 7 | YES (2) | speed | Right | NO | NO |
| 2 | AA | 19 | 6 | NO | non-speed | Right | NO | NO |
| 3 | AA | 21 | 14 | YES (1) | non-speed | Right | NO | NO |
| 4 | AA | 20 | 9 | NO | non-speed | Right | NO | NO |
| 5 | AA | 19 | - | NO | speed | Right | NO | NO |
| 6 | AA | 21 | 9 | NO | non-speed | Right | NO | NO |
| 7 | W | 21 | 12 | NO | non-speed | Right | NO | NO |
| 8 | W | 20 | 12 | YES (1) | non-speed | Right | NO | NO |
| 9 | AA | 21 | 13 | NO | non-speed | Right | NO | NO |
| 10 | W | 23 | 19 | YES (2) | non-speed | Right | NO | Dyslexia |
| 11 | AA | 21 | 11 | NO | speed | Right | NO | NO |
| 12 | AA | 19 | 8 | NO | speed | Right | NO | NO |
| 13 | W | 21 | 15 | YES (1) | non-speed | Right | NO | NO |
| 14 | W | 21 | 13 | YES (1) | non-speed | Right | NO | NO |
| 15 | W | 20 | - | NO | non-speed | Right | NO | NO |
| 16 | W | 23 | - | YES (1) | non-speed | Right | NO | NO |
| 17 | AA | 19 | 13 | YES (1) | non-speed | Right | NO | NO |
| 18 | W | 20 | 7 | NO | non-speed | Right | NO | ADHD |
| 19 | W | 23 | 14 | YES (1) | speed | Right | NO | NO |
| 20 | AA | 21 | 13 | NO | non-speed | Right | NO | NO |
| 21 | W | 20 | 6 | NO | speed | Right | NO | NO |
| 22 | AA | 23 | 8 | NO | speed | Right | NO | ADHD |
| 23 | W | 22 | 13 | NO | non-speed | Right | NO | NO |

**Supplemental Table 2**. Tables (A-J) lists all pairwise two-way associations at a significance level of 0.05 between the variables: Network *fingerprint similarity*, network-ΔFA (networks: Visual (VIS), Somatomotor (SM), Dorsal Attention (DA), Ventral Attention (VA), Limbic System (L), Fronto-Parietal (FP), Default Mode Network (DMN) and subcortical regions (SUBC)), ΔmiRNA (miR-20a, miR-505, miR-3623p, miR-30d, miR-92a, miR-486, miR-195, miR-93p, miR-151-5p),

ΔVR (Spatial Memory, Sensory-Motor reactivity (SR), Balance and Comprehensive score) and HAE measures. Cook's outliers (*k*/*n*) lists the number of outliers *k*, based on Cook's distance, removed out of the total *n* samples. The linear regression slopes β, *p*-values and Adjusted $R^2$ ($R^2_{adj}$) are reported after the outlier removal.

**(A)**

| **Network-ΔFA** | **Network *fingerprint similarity*** | **Cook's Outliers** | **β** | ***p*** | **$R^2_{adj}$** |
|---|---|---|---|---|---|
| SM-ΔFA | SM | 3/17 | 0.998 | 0.018 | 0.332 |
| L-ΔFA | L | 1/17 | 0.417 | 0.040 | 0.215 |

**(B)**

| **Network-ΔFA** | **ΔmiRNA** | **Cook's Outliers** | **β** | ***p*** | **$R^2_{adj}$** |
|---|---|---|---|---|---|
| SM-ΔFA | 20a | 3/14 | 0.824 | 0.050 | 0.293 |
| SM-ΔFA | 30d | 1/14 | 1.160 | 0.007 | 0.447 |
| VA-ΔFA | 505 | 0/14 | -0.662 | 0.010 | 0.391 |
| VA-ΔFA | 92a | 0/14 | -0.663 | 0.010 | 0.393 |
| VA-ΔFA | 195 | 1/14 | -0.990 | 0.037 | 0.279 |
| L-ΔFA | 92a | 0/14 | -0.675 | 0.008 | 0.410 |
| L-ΔFA | 486 | 0/14 | -0.667 | 0.009 | 0.399 |
| L-ΔFA | 195 | 0/14 | -0.685 | 0.007 | 0.424 |

**(C)**

| **Network-ΔFA** | **ΔVR task** | **Cook's Outliers** | **β** | ***p*** | **$R^2_{adj}$** |
|---|---|---|---|---|---|
| SM-ΔFA | ΔSR | 0/17 | -0.540 | 0.025 | 0.244 |
| DA-ΔFA | ΔBalance | 2/17 | 0.742 | 0.008 | 0.390 |
| DA-ΔFA | ΔSR | 2/17 | -0.667 | 0.017 | 0.318 |

**(D)**

| ΔVR task | Network *fingerprint similarity* | Cook's Outliers | β | *p* | $R^2_{adj}$ |
|---|---|---|---|---|---|
| ΔComprehensive | SM | 2/20 | -0.684 | 0.010 | 0.311 |
| ΔComprehensive | DA | 1/20 | -0.570 | 0.049 | 0.162 |
| ΔBalance | FC | 1/20 | 0.439 | 0.020 | 0.222 |
| ΔBalance | FP | 1/20 | 0.538 | 0.020 | 0.232 |
| ΔBalance | DMN | 1/20 | 0.447 | 0.020 | 0.240 |
| ΔBalance | SUBC | 1/20 | 0.430 | 0.030 | 0.202 |
| ΔSR | Vis | 1/20 | -0.642 | 0.030 | 0.193 |
| ΔSR | SM | 1/20 | -0.745 | 0.001 | 0.473 |
| ΔSR | DA | 1/20 | -0.702 | 0.001 | 0.466 |

**(E)**

| ΔVR task | ΔmiRNA | Cook's Outliers | β | *p* | $R^2_{adj}$ |
|---|---|---|---|---|---|
| ΔComprehensive | 505 | 1/20 | -0.671 | 0.003 | 0.387 |
| ΔComprehensive | 30d | 0/20 | -0.586 | 0.007 | 0.306 |
| ΔComprehensive | 92a | 0/20 | -0.479 | 0.033 | 0.186 |
| ΔComprehensive | 151-5p | 1/20 | -0.670 | 0.002 | 0.395 |
| ΔBalance | 505 | 1/20 | -0.533 | 0.003 | 0.389 |
| ΔBalance | 30d | 1/20 | -0.417 | 0.027 | 0.213 |
| ΔSR | 20a | 0/20 | -0.458 | 0.043 | 0.184 |
| ΔSR | 505 | 1/20 | -0.569 | 0.016 | 0.254 |
| ΔSR | 30d | 0/20 | -0.483 | 0.031 | 0.190 |
| ΔSR | 92a | 0/20 | -0.537 | 0.015 | 0.249 |
| ΔSR | 151-5p | 1/20 | -0.659 | 0.007 | 0.323 |

**(F)**

| ΔmiRNA | Network *fingerprint similarity* | Cook's Outliers | β | *p* | $R^2_{adj}$ |
|---|---|---|---|---|---|
| 505 | DMN | 1/17 | -0.458 | 0.020 | 0.302 |
| 92a | DMN | 1/17 | -0.365 | 0.015 | 0.306 |

**(G)**

| Network-ΔFA | HAE | Cook's Outliers | β | *p* | $R^2_{adj}$ |
|---|---|---|---|---|---|
| SM-ΔFA | sessions | 0/17 | 0.553 | 0.021 | 0.260 |
| SUBC-ΔFA | sessions | 1/17 | 0.752 | 0.000 | 0.601 |
| SUBC-ΔFA | $cHAE_{25G}$ | 1/17 | 0.601 | 0.028 | 0.249 |

**(H)**

| ΔmiRNA | HAE | Cook's Outliers | β | *p* | $R^2_{adj}$ |
|---|---|---|---|---|---|
| 362-3p | $cHAE_{80G}$ | 3/20 | -0.488 | 0.045 | 0.191 |
| 505 | $aHAE_{25G}$ | 1/20 | 0.397 | 0.035 | 0.191 |
| 92a | $aHAE_{80G}$ | 2/20 | 0.882 | 0.046 | 0.178 |

**(I)**

| ΔVR task | HAE | Cook's Outliers | β | *p* | $R^2_{adj}$ |
|---|---|---|---|---|---|
| ΔSR | $cHAE_{25G}$ | 1/23 | -0.582 | 0.007 | 0.278 |

**(J)**

| Network *fingerprint similarity* | HAE | Cook's Outliers | β | *p* | $R^2_{adj}$ |
|---|---|---|---|---|---|
| SM | sessions | 0/20 | 0.479 | 0.032 | 0.187 |
| DMN | $cHAE_{80G}$ | 1/20 | -0.557 | 0.031 | 0.201 |
| SUBC | $cHAE_{80G}$ | 1/20 | -0.692 | 0.001 | 0.452 |
| SUBC | $aHAE_{25G}$ | 1/20 | -0.609 | 0.009 | 0.302 |
| SUBC | $aHAE_{80G}$ | 1/20 | -0.992 | 0.001 | 0.432 |
| FP | $aHAE_{25G}$ | 2/20 | -0.519 | 0.026 | 0.227 |
| VIS | $aHAE_{80G}$ | 1/20 | -0.401 | 0.019 | 0.242 |

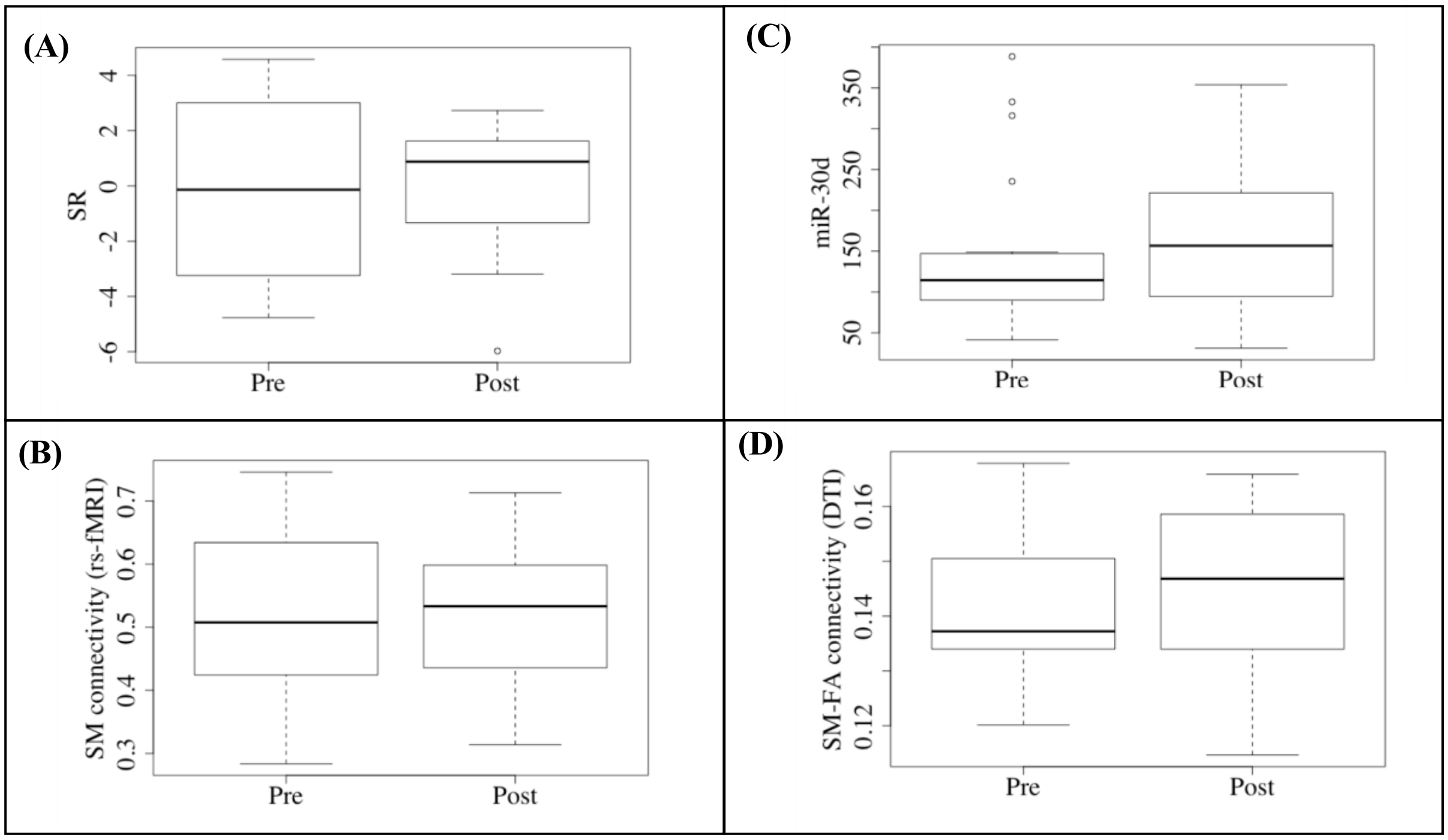

**Supplemental Figure 1.** Boxplots comparing the *Pre*- and *Post*-season distributions of (A) Sensory-Motor reactivity (SR) (B) Somatomotor (SM) connectivity (rs-fMRI) (C) miR-30d and (D) Somatomotor (SM) average FA connectivity (DTI). Note that only miR-30d significantly increased from *Pre* to *Post*-season using paired *t*-test.

## Code for Permutation-based Mediation in R

```
# Permutation based Mediations for SC+FC+VR : Post-Pre
rm(list = ls())     # clear objects
graphics.off()      # close graphics windows
# Last Edited : 01/21/2022 by Sumra Bari
# Load the data -----------------------------------------------------
# set the working directory
setwd('/Users/SumraBari/Desktop/PSU/DTI-omics/Scientific Reports/to_upload')

path_to_folder=getwd()
path_to_file=file.path(path_to_folder,'PrePost_VR_miRNA.csv',fsep = '/')
mydata=read.csv(path_to_file,header = TRUE ,sep = ',')

path_to_file=file.path(path_to_folder,'PrePost_DTI_SM.csv',fsep = '/')
DTI=read.csv(path_to_file,header = FALSE ,sep = ',')

path_to_file=file.path(path_to_folder,'PrePost_rsfMRI_SM.csv',fsep = '/')
rsfMRI=read.csv(path_to_file,header = FALSE ,sep = ',')

# The common subjects to keep------------------------------------------
# subjects with complete data
tokeep = c( 1, 2, 3, 4, 5, 6, 8, 9, 10, 11, 13, 15, 16, 19, 20, 22, 23)
# make the not common rows to NA
mydata[!is.element(mydata$Subject.ID,tokeep),]<-NA
# delete the rows with all NAs -----------------------------
mydata = mydata[rowSums(is.na(mydata)) != ncol(mydata),]
# keep the common subjects for DTI
DTI = DTI[,is.element(DTI[1,],tokeep)]
# remove the subject ID
DTI = DTI[-1,]
# keep the common subjects for rsfMRI
rsfMRI = rsfMRI[,is.element(rsfMRI[1,],tokeep)]
# remove the subject ID
rsfMRI = rsfMRI[-1,]
# Set the number of Permutations -------------------------------------
Q = 100000
S = length(tokeep) # number of subjects
N = 2*S  # total time-points

# Assign values to the variables--------------------------------------
xdata = mydata$VR.SR
mdata = DTI
ydata = rsfMRI


x1 = xdata[mydata$Session==0]
m1 = DTI[,1:S]
y1 = rsfMRI[,1:S]

x2 = xdata[mydata$Session==1]
m2 = DTI[,(S+1):(2*S)]
y2 = rsfMRI[,(S+1):(2*S)]
```

```
# Get z0 without permutation ---------------------------------

xdelta = x2-x1 #calculate the Post-Pre
mdelta = sqrt(colSums((m2 - m1)^2)) #delta FA across network
ydelta = diag(cor(y1,y2)) # fingerprint similarity

# step1: y~x
fit1 = lm (ydelta ~ xdelta)
# find the cook's distance and the outliers which exceed 4/n
bad1 = cooks.distance(fit1) > 4/length(xdelta)

# step2: m ~x # change
fit2 = lm (mdelta ~ xdelta)
bad2 = cooks.distance(fit2) > 4/length(xdelta)

# step3: y ~ m
fit3 = lm (ydelta~mdelta)
bad3 = cooks.distance(fit3) > 4/length(mdelta)

# step4 : y ~ m+x
# remove all the outliers from step1 and step2
bad = bad1 | bad2 | bad3
x=xdelta[!bad]
m=mdelta[!bad]
y=ydelta[!bad]

# Sobel Test
library(multilevel)
#change
s = sobel(x,m,y)
# sobel effect size
# indirect effect/total effect
sobel_eff = (s$Indirect.Effect/s$`Mod1: Y~X`[2,1])*100
# sobel p-value
sobel_p= 2*pnorm(abs(s$z.value),lower.tail = FALSE)
z0 = s$z.value
# print out the sobel-p and sobel-eff
cat('Without Permutations \n')
cat('Sobel-p = ',sobel_p,'\n')
cat('Sobel-Effect = ', sobel_eff,'\n')
# Permute data and run Moderations ----------------------------------

zq = rep (0,Q)

for (q in 1:Q) {

   ind1 = sample(c(1:N),S)
   ind2 = setdiff(c(1:N),ind1)
   x1q = xdata[ind1]
   x2q = xdata[ind2]
   xdelta = x2-x1 #calculate the Post-Pre

   ind1 = sample(c(1:N),S)
   ind2 = setdiff(c(1:N),ind1)
   m1q = DTI[,ind1]
   m2q = DTI[,ind2]
```

```
    mdelta = sqrt(colSums((m2q - m1q)^2))  #delta FA across network

    ind1 = sample(c(1:N),S)
    ind2 = setdiff(c(1:N),ind1)
    y1q = rsfMRI[,ind1]
    y2q = rsfMRI[,ind2]
    ydelta = diag(cor(y1q, y2q )) # fingerprint similarity


    # step1: y~x
    fit1 = lm (ydelta ~ xdelta)
    # find the cook's distance and the outliers which exceed 4/n
    bad1 = cooks.distance(fit1) > 4/length(xdelta)

    # step2: m ~x
    #change
    fit2 = lm (mdelta ~ xdelta)
    bad2 = cooks.distance(fit2) > 4/length(xdelta)

    # step3: y ~ m
    fit3 = lm (ydelta~mdelta)
    bad3 = cooks.distance(fit3) > 4/length(mdelta)

    # step4 : y ~ m+x
    # remove all the outliers from step1 and step2
    bad = bad1 | bad2 | bad3
    x=xdelta[!bad]
    m=mdelta[!bad]
    y=ydelta[!bad]

    # Sobel Test
    library(multilevel)
    #change
    s = sobel(x,m,y)
    zq[q] = s$z.value
}

# Get the p-value for beta3 and F-stat ----------------------------
pz = (sum (abs(zq)>abs(z0)))/ Q
# Print out the results
cat('Permutation based p-values \n')
cat('Sobel-p-perm = ',pz,'\n')
```

## Code for Permutation-based Moderation in R

```
# Permutation based Moderations: DTI-omics

rm(list = ls())     # clear objects

graphics.off()      # close graphics windows

# Last Edited : 01/21/2022 by Sumra Bari


# Load the data -------------------------------------------------------

# set the working directory

setwd('/Users/SumraBari/Desktop/PSU/DTI-omics/Scientific Reports/to_upload')

path_to_folder=getwd()

path_to_file=file.path(path_to_folder,'PrePost_VR_miRNA.csv',fsep = '/')

mydata=read.csv(path_to_file,header = TRUE ,sep = ',')

path_to_file=file.path(path_to_folder,'PrePost_DTI_SM.csv',fsep = '/')

DTI=read.csv(path_to_file,header = FALSE ,sep = ',')


# The common subjects to keep-----------------------------------------

# subjects with complete data

tokeep = c( 1, 2, 3, 4, 5, 6, 8, 9, 10, 11, 13, 15, 16, 20)

# make the not common rows to NA

mydata[!is.element(mydata$Subject.ID,tokeep),]<-NA

# delete the rows with all NAs

mydata = mydata[rowSums(is.na(mydata)) != ncol(mydata),]

# keep the common subjects for DTI

DTI = DTI[,is.element(DTI[1,],tokeep)]

# remove the subject ID

DTI = DTI[-1,]

# Set the number of Permutations -------------------------------------

Q = 100000
```

```
S = length(tokeep) # number of subjects

N = 2*S  # total time-points

# Assign values to the variables--------------------------------------

xdata = DTI

mdata = mydata$VR.SR

ydata = mydata$miR.30d


x1 = DTI[,1:S]

m1 = mydata$VR.SR[mydata$Session == 0]

y1 = mydata$miR.30d[mydata$Session == 0]


x2 = DTI[,(S+1):(2*S)]

m2 = mydata$VR.SR[mydata$Session == 1]

y2 = mydata$miR.30d[mydata$Session == 1]


# Get t0 and F0 without permutation ----------------------------------

xdelta = sqrt(colSums((x2 - x1)^2))

ydelta = y2-y1

mdelta = m2-m1

# fit the data

fit= lm(ydelta~ xdelta+ mdelta + xdelta*mdelta)

# get cook's distance and remove outliers

bad = cooks.distance(fit) > 4/length(xdelta)

y=ydelta[!bad]

m=mdelta[!bad]

x=xdelta[!bad]

# run regression again

fit1 = lm (y~ x+ m + x*m)
```

```
t0 = summary(fit1)$coefficients[4,3]

F0 = summary(fit1)$fstatistic[1]

# print p-values

cat('Without Permutations \n')

cat('pbeta3 = ',summary(fit1)$coefficients[4,4],'\n')

pF = pf(summary(fit1)$fstatistic[1],summary(fit1)$fstatistic[2],

        summary(fit1)$fstatistic[3],lower.tail = FALSE)

cat('pF = ',pF,'\n')

# Permute data and run Moderations ----------------------------------


tq = rep (0,Q)

Fq = rep (0,Q)

for (q in 1:Q) {


    ind1 = sample(c(1:N),S)

    ind2 = setdiff(c(1:N),ind1)

    x1q = DTI[,ind1]

    x2q = DTI[,ind2]

    xdelta = sqrt(colSums((x2q - x1q)^2))


    ind1 = sample(c(1:N),S)

    ind2 = setdiff(c(1:N),ind1)

    y1q = ydata[ind1]

    y2q = ydata[ind2]

    ydelta = y2q-y1q


    ind1 = sample(c(1:N),S)

    ind2 = setdiff(c(1:N),ind1)
```

```
    m1q = mdata[ind1]

    m2q = mdata[ind2]

    mdelta = m2q-m1q


    # fit the data

    fit= lm(ydelta~ xdelta+ mdelta + xdelta*mdelta)

    # get cook's distance and remove outliers

    bad = cooks.distance(fit) > 4/length(xdelta)

    x=xdelta[!bad]

    m=mdelta[!bad]

    y=ydelta[!bad]

    # run regression again

    fit1 = lm (y~ x+ m + x*m)

    tq[q] = summary(fit1)$coefficients[4,3]

    Fq[q] = summary(fit1)$fstatistic[1]

}


# Get the p-value for beta3 and F-stat ----------------------------

pbeta3 = (sum (abs(tq)>abs(t0)))/ Q

pF = (sum(Fq > F0))/Q

# Print out the results

cat('Permutation based p-values \n')

cat('pbeta3 = ',pbeta3,'\n')

cat('pF = ',pF,'\n')
```